\documentstyle[12pt]{article}
\newcommand{\sbody}[2]{{\textstyle\frac{#1}{#2}}}
\newcommand{\stk}[1]{\stackrel{*}{\overline}}

\begin{document}
\begin{center}
\vfill
\large\bf{$AdS_2$ Models in an Embedding Superspace }
\end{center}
\begin{center}
D.G.C. McKeon$^{(1)}$\\
T.N. Sherry$^{(2)}$\vspace{.3cm}\\
Department of Mathematics/Science\\
State University of New York, Utica/Rome\\
Utica, New York, U.S.A.\\
13504-3050\vspace{.6cm}\\
\end{center}
\begin{tabbing}
$^{(1)}$ Permanent Address: \= Department of Applied Mathematics\vspace{-.3cm} \\
	   		    \> University of Western Ontario\vspace{-.3cm}\\
			    \> London  CANADA \vspace{-.3cm}\\
			    \> N6A 5B7\vspace{-.3cm}\\
	             \= email: DGMCKEO2@UWO.CA
\end{tabbing}
\begin{tabbing}
$^{(2)}$ Permanent Address: \= Department of Mathematical Physics\vspace{-.3cm} \\
	   		    \> National University of Ireland, Galway\vspace{-.3cm}\\
			    \> Galway  IRELAND \vspace{-.3cm}\\
	             \= email: TOM.SHERRY@NUIGALWAY.IE
\end{tabbing} 

\noindent{\Large\bf{Abstract}}

An embedding superspace, whose Bosonic part is the flat $2 + 1$ dimensional embedding
space for $AdS_2$, is introduced.  Superfields and several supersymmetric models
are examined in the embedded $AdS_2$ superspace\vspace{.6cm}.
\eject

The N = 1 supersymmetry (SUSY) algebra in $AdS_2$ space can be written\footnote{The
metric is $\eta_{\mu\nu} = \rm{diag} (+, -, +)$ and we take $\gamma^\mu\gamma^\nu
= -\eta^{\mu\nu} + 2\Sigma^{\mu\nu}$ with $\gamma_1 = 
i\tau_1,\,\gamma_2 = \tau_2,\, \gamma_3 = i\tau_3$ so that 
$\Sigma^{\mu\nu} L_{\mu\nu} \Sigma^{\lambda\sigma} L_{\lambda\sigma} - 
\Sigma^{\mu\nu} L_{\mu\nu} = -\frac{1}{2} L^{\mu\nu} L_{\mu\nu}$
.  $Q$ is Majorana with $\overline{Q} = Q^\dagger
\gamma_2$, $\tilde{Q} = Q^T\gamma_2$, $Q = Q_C = \gamma_2\overline{Q}^T$.}
$$\left[J_{\mu\nu}, J_{\lambda\sigma} \right] = \eta_{\mu\lambda} J_{\nu\sigma} -
\eta_{\nu\lambda} J_{\mu\sigma} + \eta_{\nu\sigma} J_{\mu\lambda} - \eta_{\mu\sigma}
J_{\nu\lambda}\eqno(1a)$$
$$\left[J_{\mu\nu}, Q\right] = -\Sigma_{\mu\nu} Q\eqno(1b)$$
$$\left\lbrace Q, \tilde{Q} \right\rbrace = 2 \Sigma_{\mu\nu} J^{\mu\nu} .\eqno(1c)$$
$AdS_2$ corresponds to a surface of constant curvature embedded in flat 2 + 1 
dimensional space time (2+1D) described by the constraint
$$x^\mu x_\mu = a^2\; .\eqno(2)$$

In this letter, we follow Dirac's approach to $dS_4$ space [1]; 
we work in the embedding space, rather than
directly in the two dimensional curved space itself, and we
use the $S0(2,1)$ generators associated with the symmetries on this curved
surface.  A similar approach has been used by Adler [3] and 
Drummond and Shore [4] to analyze massless gauge
theories on the surface $S_4$, using the $S0(5)$ generators 
associated with $S_4$.

In a previous paper [2] we constructed, using
this approach, a component field supersymmetric model involving two real spin zero
fields and one Majorana spinor field. These fields we have shown to carry full off
shell implementation of the SUSY algebra (1) above.

In this letter we wish to report on some progress in developing a
superspace and superfield method of addressing this problem. Our
approach is modelled directly on Dirac's approach.  What we do in this paper
is distinct from the approach employed in [5,6] where
$AdS_2$ supersymmetry is 
implemented directly in $2|2$ curved superspace [7,8,9] by specializing to
the appropriate fixed geometry. 

We consider a
superspace with a 2 + 1 D Bosonic part, corresponding to the embedding
space for $AdS_2$, with coordinates $x^\mu$ $(\mu = 1, 2, 3)$, and a
Fermionic part with Grassmann coordinates given by a Majorana spinor
$\theta_i (i = 1, 2)$. The $AdS_2$ superspace is an embedded surface in
this superspace given by the supersymmetric generalization of the
above constraint.

In this embedding superspace we can realise the generators in the
algebra (1) as follows
$$J_{\mu\nu} = \frac{\partial}{\partial\theta} \Sigma_{\mu\nu} \theta -
\left(x_\mu \partial_\nu - x_\nu\partial_\mu\right) \equiv
\frac{\partial}{\partial\theta} \Sigma_{\mu\nu} \theta +
L_{\mu\nu}\eqno(3)$$
$$Q = \gamma^\mu\partial_\mu\theta + \gamma^\mu x_\mu
\frac{\partial}{\partial\tilde{\theta}}\eqno(4)$$
$$\tilde{Q} = - \tilde{\theta} \gamma^\mu\partial_\mu +
\frac{\partial}{\partial\theta} \gamma^\mu x_\mu \;\; .\eqno(5)$$

(We do not include the translation operator in the Bosonic
sector of this algebra, as in [13].)

We can identify at once the supersymmetric generalization of two SO(2,1)
invariants which are crucial to Dirac's approach. First, the
generalization of the SO(2,1) invariant $x^\mu x_\mu$ is
$$R^2 \equiv x^\mu x_\mu - \tilde{\theta}\theta\eqno(6)$$
and second, the generalization of the SO(2,1) invariant $x^\mu
\partial_\mu$ is
$$\Delta \equiv x^\mu \partial_\mu + \theta_i
\,\frac{\partial}{\partial\theta_i} = x^\mu \partial_\mu +
\tilde{\theta}_i\,\frac{\partial}{\partial\tilde{\theta}_i}\;
.\eqno(7)$$
In other words, both $R^2$ and $\Delta$ commute with $J_{\mu\nu}$ and
$Q$. Dirac [1] used the constraint $x^\mu x_\mu = a^2$ to go from the
embedding space to $dS_4$ space; we
use its supersymmetric generalization to go from the embedding superspace 
to the $AdS_2$ superspace. This entails including a factor of $\delta\left(R^2
- a^2\right) = \delta\left(x^2 - a^2\right) - \tilde{\theta}\theta\delta^{\prime}
\left(x - a^2\right)$ in the action.
Also Dirac 
used $x^\mu \partial_\mu$ to define fields, initially given on the $dS_4$
surface, away from that surface; likewise we use $\Delta$ to define
superfields, initially given on the $AdS_2$ supersurface, away from that
surface.

A scalar superfield will be a scalar function defined on the embedding
superspace, $\Phi(x,\theta)$. In the Dirac approach, the superfield is
considered defined on the surface $x^2 = a^2$ by including the supersymmetry
invariant factor of $\delta\left(R^2 - a^2\right)$ in the action.
It is defined away from this
surface by imposing the condition
$$\Delta\Phi = w\Phi\eqno(8)$$
where $w$ is a real eigenvalue. We note that this condition is preserved
under a supersymmetry transformation. Expanding $\Phi$ in powers of
$\theta$ 
$$\Phi(x,\theta) = \phi (x) + \tilde{\lambda} (x)\theta + F(x)
\tilde{\theta}\theta \; .\eqno(9)$$
we can at once identify the component fields, namely, 
two scalar fields $\phi$ and $F$ and a Majorana spinor field $\lambda$.
The condition (8) translates into the following homogeneity conditions
for the component fields
$$\left(x^\mu \partial_\mu - w\right)\phi = \left(x^\mu \partial_\mu + 1 - w\right)
\lambda = \left(x^\mu \partial_\mu + 2 - w\right)F = 0\; .\eqno(10)$$
The SUSY transformations for the component fields are induced from
$$\delta\Phi = i\left[\tilde{\xi} Q, \Phi\right]\eqno(11)$$
by the realization (4); we find
$$\delta\phi = i\tilde{\xi}\gamma^\mu x_\mu\lambda\eqno(12a)$$
$$\delta\tilde{\lambda} = i\tilde{\xi}\gamma^\mu\left(\partial_\mu \phi + 2x_\mu
F\right)\eqno(12b)$$
$$\delta F = -\sbody{i}{2}\tilde{\xi} \gamma\cdot \partial \lambda \;
.\eqno(12c)$$
It is now possible to specify an action in terms of the superfield $\Phi$
that is automatically invariant under the SUSY transformations associated
with the $AdS_2$ algebra (1).  To do this we introduce the operators $E(\alpha , \beta)$
defined by
$$E(\alpha , \beta) = \alpha \gamma \cdot \partial \theta + 
\beta \gamma \cdot x \frac{\partial}{\partial \tilde{\theta}}\eqno(13a)$$
$${\tilde{E}}(\alpha , \beta) = -\alpha {\tilde{\theta}} \gamma \cdot \partial + 
\beta \frac{\partial}{\partial \theta}\gamma \cdot x .\eqno(13b)$$
If $\alpha = 1/\beta$ (and only in this case), $E(1/\beta , \beta )\,({\tilde{=}} Q)$
satisfies the algebra of eq. (1c).
Furthermore, 
$$[E(\alpha , \beta), \Delta ] = 0\eqno(14a)$$
for all $\alpha$, $\beta$ while
$$[E(\alpha , \beta), R^2 ] \neq 0\eqno(14b)$$
unless $\alpha = \beta$. It is now possible to write down a family of free
supersymmetric actions for the superfield $\Phi$ in the embedding $AdS_2$ superspace,
namely
$$S_1(\alpha , \beta , \rho) = \int d^3x\, d^2\theta\delta\!\left(R^2 - a^2\right)\Phi
\left(\tilde{E} E + \rho\right)\Phi\; .\eqno(15)$$

The role of the $\delta$-function is to constrain the integration over the
embedding superspace to just the $AdS_2$ superspace part of it. It implies that
the Bosonic part of the superspace is constrained to lie on the $AdS_2$ surface
embedded in $2 + 1$ dimensions as
$$\delta\!\left(R^2 - a^2\right) = \delta\!\left(x^2 - a^2\right) -
\delta^\prime\!\left(x^2 - a^2\right)\tilde{\theta}\theta \,\, .\eqno(16)$$

Clearly $S_1$
is SO(2,1) invariant; it is also supersymmetric. (The product of the
three scalar superfields $\Phi$, $\tilde{E} E\Phi$ and $\delta\left(R^2
- a^2\right)$ is also a scalar superfield, and by (12c) the
$\tilde{\theta}\theta$ component of any scalar superfield transforms as a
divergence. Upon defining $\int d^2\theta\,\tilde{\theta}\,\theta = 1$,
the $\theta$ integration in (15) picks out the $\tilde{\theta}\theta$
component of the superfield integrand, and thus it follows that (15) is
supersymmetric for all values of $\alpha$, $\beta$, and $\rho$.)

The component field form of this action can be deduced using
$$\tilde{E}E = -\beta^2x^2 \frac{\partial}{\partial\theta}
\frac{\partial}{\partial\tilde{\theta}} + \frac{\alpha^2}{2x^2}
\left(L^{\mu\nu} L_{\mu\nu} + 2 (x \cdot \partial )^2 + 2(x \cdot
\partial)\right)\tilde{\theta}\theta\eqno(17a)$$
$$-\alpha \beta \left[ 2 (x \cdot \partial ) + 2\tilde{\theta}\left(-x\cdot\partial +
\Sigma^{\mu\nu}L_{\mu\nu}\right)\,\frac{\partial}{\partial\tilde{\theta}}
- 3 \tilde{\theta} \frac{\theta}{\partial\tilde{\theta}}\right]. \nonumber$$
We find that
$$S_1(\alpha , \beta , \rho) = \int d^3x \left\lbrace \delta\!\left( x^2 - a^2\right)
\left[ +\tilde{\lambda}\left(\alpha\beta\Sigma^{\mu\nu} L_{\mu\nu}-\frac{\rho +
3\alpha\beta}{2}\right)\lambda + \frac{\alpha^2}{2x^2}
\phi\left(L^{\mu\nu}L_{\mu\nu}
+ 2w(1 + w)\right)\phi\right.\right.\eqno(17b)$$
$$\left.\left. - 2x^2 \beta^2 F^2 + 2(\rho + \alpha\beta)\phi
F\right] + \delta^\prime \left(x^2 - a^2\right)\left[ 2x^2 \beta^2 \phi F - (\rho - 2\alpha
\beta\omega)
\phi^2 \right]\right\rbrace\; .\nonumber$$
Here the action is expressed as an integral over the embedding space of
$AdS_2$. Integrating over $\sqrt{x^\mu x_\mu}$ using the 
$\delta$-function\footnote{We use 
$$\int d^3x\, \delta^\prime \left(x^2 - a^2\right)f(x) =
-\frac{1}{2a^2} \int d^3x \,\delta\left(x^2 - a^2\right)(x \cdot \partial + 1) f(x)
\nonumber$$
$$\;\;\;\;\;\;\;\;\;\left. \equiv -\frac{1}{2a^2} \int d^2A 
\frac{a}{2} 
(x \cdot \partial + 1) f(x)\right|_{\left(x^2 = a^2\right)}
\nonumber$$}
we find the action integral defined on the $AdS_2$
surface
$$S_1(\alpha ,\beta ,\rho) = \int d^2A \left(\frac{a}{2}\right)
\left\lbrace {\tilde{\lambda}} \left[ \alpha\beta \left(\Sigma^{\mu\nu}
L_{\mu\nu} - \frac{3}{2}\right) - \rho /2 \right]\lambda\right.\eqno(18)$$
$$+ \frac{1}{2a^2} \phi \left[\alpha^2\left(L^{\mu\nu} L_{\mu\nu}
+ 2w (1 + w)\right) + (2w + 1)(\rho - 2\alpha\beta w)\right]\phi\nonumber$$
$$\left. -2 a^2 \beta^2 F^2 + \left[ -\beta^2 (2w+1) + 2 (\rho +
\alpha \beta)\right]\phi F\right\rbrace .\nonumber$$

It is of interest to compare this supersymmetric action on $AdS_2$
derived by superfield methods with the supersymmetric action
constructed previously [2] without the benefit of superfield techniques.
The free part of that action was (with $a^2 = 1$), 
$${\overline{S}} = \int d^2 A \left[- \tilde{\Psi} \left(\Sigma^{\mu\nu} L_{\mu\nu}
+ \chi + \lambda_1\right) \Psi + A\left(\frac{1}{2} L^{\mu\nu}
L_{\mu\nu}\right.\right.\eqno(19)$$
$$ + \chi(1 + \chi )+ \lambda_1 (1 + 2\chi )\biggr)A - B^2 +
2\lambda_1 AB\biggr],\nonumber$$
where $A$ and $B$ are real scalar fields and $\Psi$ is a Majorana
spinor. The SUSY transformations under which $\overline{S}$ is invariant are
$$\delta A = \tilde{\xi}\Psi\eqno(20a)$$
$$\delta \Psi = -\left[\left(\Sigma^{\mu\nu} L_{\mu\nu} - (1 +
\chi)\right)A - B\right]\xi\eqno(20b)$$
$$\delta B = -\tilde{\xi} \left[\Sigma^{\mu\nu} L_{\mu\nu} + \chi\right]\Psi\; .\eqno(20c)$$
These transformations satisfy the algebra of eq. (1c).

Let us compare the two actions (18) and (19) and the SUSY
transformations under which they are invariant, (12) and (20)
respectively. We note that the transformations (12) and (20) are identical if we
identify $A = \phi$, $\Psi = i\gamma \cdot x\lambda$, $B = -2x^2 F$ and
$\chi = - 1 - w$.  However, the action $\overline{S}$ in (19) does not yield $S_1$ in (18)
if we make this identification;
we find (using $\Sigma^{\mu\nu} L_{\mu\nu} \gamma \cdot x =
\gamma \cdot x \Sigma^{\mu\nu} L_{\mu\nu} -2 \gamma \cdot x$)
$$\overline{S} = \int d^2 A \left\lbrace {\tilde{\lambda}} \left(\Sigma^{\mu\nu}
L_{\mu\nu} - 2 + \chi + \lambda_1\right)\lambda\right. \eqno(21)$$
$$+ \phi \left(\frac{1}{2} L^{\mu\nu} L_{\mu\nu} + \chi (1 + \chi) + \lambda_1
(1 + 2\chi)\right)\phi\nonumber$$
$$\left. -4F^2 - 4\lambda_1 \phi F\right\rbrace\, .\nonumber$$
This form of $\overline{S}$ and the form of $S_1$ in (18) cannot be equated
for any values of $\alpha$, $\beta$ and $\rho$.  However, both are invariant under the
same SUSY transformations, namely eqs. (12).

It is possible to write down another family of supersymmetric actions in 
superfield formalism, 
$$S_2(\alpha , \beta ,\rho) = \int d^3x d^2\theta \delta\left(R^2 - a^2\right)
\left[-\left({\tilde{E}}\Phi\right)\left(E\Phi\right) + \rho\Phi^2\right].\eqno(22)$$
The actions $S_2$ are distinct from the actions of $S_1$ unless
$[E,R^2] = 0$, which only occurs when $\alpha = \beta$. $S_2$ is clearly supersymmetric;
the product of the spinor superfields ${\tilde{E}}\Phi$ and $E\Phi$ is a scalar superfield
and hence the integrand is a scalar superfield whose $F$ component transform
as a total divergence (12c).

The actions of eqs. (15) and (19) do not involve interactions. A suitable
interaction in the superfield approach is given by
$$S_N = g_N \int d^3x d^2\theta \delta\left(R^2 - a^2\right)\Phi^N\;\;\;(N \geq 3).
\eqno(23)$$
It is also possible to supplement the action $\overline{S}$ of eq. (19)
with an interaction Lagrangian
$${\overline{S}}_N = 
\lambda_N \int d^3x \left[(1 + 2\chi)A^2 - {\tilde{\Psi}}\Psi +
2AB\right]^N
\eqno(24)$$
$(N \geq 2)$. For all $N$, ${\overline{S}}_N$ in (24) is invariant under
the transformations of eq. (20). The interactions $S_N$
and ${\overline{S}}_N$ are clearly distinct, as ${\overline{S}}_N$ contains
a contribution of order $(AB)^N$ while $S_N$ is at most linear in $F$.

The N = 2 SUSY algebra on $AdS_2$ involves replacing (1a) by
$$\left\lbrace Q_i , \tilde{Q}_j \right\rbrace = -i\epsilon_{ij} \alpha
\pm 2\delta_{ij} \Sigma^{\mu\nu} J_{\mu\nu}\eqno(25a)$$
and introducing
$$\left[\alpha , Q_i\right] = \pm i\epsilon_{ij}\alpha\;\; .\eqno(25b)$$
(Either sign is consistent with the Jacobi identities.)  The occurrence
of an `internal symmetry' generator $\alpha$ in (25) is reminiscent
of the $Z$ generator in
the simplest SUSY algebras on the surface of a sphere $S_2$
embedded in three dimensions. On $S_2$ one can only have Dirac or symplectic
Majorana (but not
Majorana) spinors. These SUSY algebras on $S_2$ are
$$\left\lbrace R, R^\dagger \right\rbrace = Z \mp 2 \vec{\sigma}
\cdot \vec{J} \;\;\;\;\; \left[ J_a, R\right] = -\frac{1}{2} \sigma_a R\eqno(26a,b)$$
$$\left[ Z, R \right] = \mp R\;\;\;\;\; \left[ J_a, J_b\right] = i\epsilon_{abc} J_c\eqno(26c,d)$$
where $R$ is a Dirac spinorial generator. A multiplet consisting of a
Dirac spinor and two complex scalars has a supersymmetric action given in [2] which is
similar in form to the sum of (19) and (24). 
It is interesting to speculate that a superfield formulation,
which carries a representation of
either of the algebras of (26), can be found for
the case of $S_2$.  Indeed, it is possible to realize the operators occurring
in (26), namely
$R$, $R^\dagger$, $Z$ and $J$, on a $4|4$ space
$$R = \left( \vec{\sigma} \cdot \vec{x} + \beta \right)
\frac{\partial}{\partial\theta^\dagger} \pm
\left(\frac{\partial}{\partial\beta} - \vec{\sigma} \cdot
\vec{\nabla}\right)\theta\;\;\;\;R^\dagger = \frac{\partial}{\partial\theta} \left( 
\vec{\sigma} \cdot \vec{x} + \beta \right)
\mp \theta^\dagger
\left(\frac{\partial}{\partial\beta} - \vec{\sigma} \cdot
\vec{\nabla}\right)
\eqno(27a,b)$$
$$J_a = \frac{1}{2} \left[ \theta^\dagger \sigma_a
\frac{\partial}{\partial\theta^\dagger} +
\frac{\partial}{\partial\theta} \sigma_a \theta \right] + \left( -i
\vec{x} \times\vec{\nabla}\right)_a\;\;\;\;Z = \pm \left[ \theta^\dagger \frac{\partial}{\partial\theta^\dagger}
- \theta \frac{\partial}{\partial\theta}\right]
\eqno(27c,d)$$
To achieve this realization we have introduced an auxiliary variable $\beta$
which does not seem to have anything to do with the sphere $S_2$.
The $S0(3)$ invariants $x^2$ and $\vec{x} \cdot \vec{\nabla}$ can be
generalized to $x^2 - \beta^2 - 2\theta^\dagger \theta$ and $\displaystyle{\vec{x} \cdot
\vec{\nabla} + \beta\frac{\partial}{\partial\beta} +
\theta^\dagger \frac{\partial}{\partial\theta^\dagger} + \theta
\frac{\partial}{\partial\theta}}$, both of which
commute with $R$, $R^\dagger$, $\vec{J}$ and $Z$.  However, no such
superfield formulation has yet been found. Indeed, it may be necessary to consider
harmonic superspace, as has been
done for $N = 2$ supersymmetry in four dimensional Minkowski space [10],
to achieve this formulation on $S_2$.

In this paper and in our previous work [2], we have chosen to follow
the approach of Dirac [1] and work in the flat embedding space for
$AdS_2$. A more conventional approach has been used to construct supersymmetric
models in $AdS_2$ [5,6].  In this approach one works directly in the two dimensional
constant curvature $AdS_2$ space.  Although a superfield formalism in $2|2$
dimensions using supergravity techniques is known [7-9], no superfield version of
the $AdS_2$ models of [5,6] has been given explicitly.

It remains to be seen what precise relationship (if any) there is
between the models which we have presented in this paper and the models
of [5,6].  
A somewhat similar problem for the spinor field has
been considered some time ago [11].  In that case the relationship between
Dirac's $dS_4$ wave equation and the spinor wave equation in a four
dimensional space of the appropriate constant curvature was examined.  
We establish below, and in the appendix, the mappings between the Bosonic and spinor free
field equations of motion in the two formulations.

We can
parametrize $AdS_2$ space using coordinates $t$ and $\rho$ where
$$x^1 = a\sin t \sec \rho\;\;\;\;\;
x^2 = a \tan\rho\;\;\;\;\;
x^3 =-a\cos t \sec\rho .
\eqno(28a)$$
This implies that in $AdS_2$ space
$$g_{11} = -g_{22} = \frac{a^2}{\cos^2\rho}\;,\;\;\;\;  w_1^{12} = -\tan\rho\;,\;\;\;\;
w_2^{12} = 0\eqno(29b,c,d)$$
while the $S0(2,1)$ generators are
$$L_{21} = \cos t\sin\rho\, \partial_t + \sin t \cos\rho\; \partial_\rho ,\eqno(29e)$$
$$L_{31} = \partial_t\;\;\;\;\;\;L_{32} = -\sin \rho\sin t\, \partial_t +
\cos t\cos\rho \,\partial_\rho .\nonumber$$
It follows that
$$\frac{1}{2} L^{\mu\nu} L_{\mu\nu} = \cos^2\rho\left(\partial^2_t -
\partial_\rho^2\right) = \frac{1}{a^2}\, g^{ab} \partial_a \partial_b
\eqno(30)$$
showing that the Bosonic field equations of motion are
equivalent in the two formulations. 
In the models of [5,6] the spinor equation of
motion is, using our conventions,
$$0 = \left(\gamma^a D_a - m\right)\chi = \left[\cos\rho\left(\gamma^1\partial_t
+ \gamma^2\partial_\rho + \gamma^1\Sigma_{12} \tan\rho\right)
-m\right]\chi \;\; . \eqno(31a)$$
In the models we have considered, in this paper, the spinor equation is of the form
$$0 = \left(\Sigma^{\mu\nu} L_{\mu\nu} - M\right)\lambda =
\left\lbrace i \left[\gamma^3\left(\cos t\sin\rho\,\partial_t +
\sin t\cos\rho\,\partial_\rho\right)\right.\right.\nonumber$$
$$\left.\left. + \gamma^2 \left(\partial_t\right) + \gamma^1\left(-\sin\rho \sin t\,\partial_t
+ \cos t\cos\rho\,\partial_\rho\right)\right] - M\right\rbrace\lambda\; .\eqno(31b)$$
Although equations (31a) and (31b) appear to be different, they are
actually equivalent.  We can provide a linear mapping between their
solutions $\chi$ and $\lambda$.  The details of this mapping are provided in the
appendix.

The whole question of quantization and the computation of radiative
effects requires examination. Some preliminary work, in the context of
the multiplet approach discussed above, has been reported in [2]. With
the development of superfield versions of models on $AdS_2$, the 
possibility of quantization in superspace and of doing computations
using superspace Feynman rules arises.

The models presented in this paper may well allow the construction of
supersymmetric non-linear sigma models in which the world sheet is an
$AdS_2$ space and the target space is a general $N$ dimensional space.
[12]

Supersymmetric models on higher dimensional spaces of constant curvature
merit consideration.

\noindent{\Large\bf{Acknowledgements}}

We would like to thank the National Science and Engineering Research
Council of Canada, the Enterprise Ireland International Collaboration
Program 2001 and the NUI Galway Millenium Fund 2000 for financial
support, as well as the KEK Theory Division, the Physics Department
at Ochanomizu University  for their hospitality where this work was 
initiated and the Department of Mathematics/Science
at SUNYIT (Utica/Rome) where this work was completed. R. and D. MacKenzie had useful suggestions.

\noindent{\Large\bf{Appendix}}

There is a linear relationship $\chi = T\lambda$ between the wave
functions $\chi$ and $\lambda$ appearing in (31a) and (31b) respectively
provided
$$\left( \gamma^a D_a - m\right) T\lambda = kT\left(\Sigma^{\mu\nu} L_{\mu\nu}
- M\right) \lambda \;.\eqno(A.1)$$
Here $k$ is a constant and
$$T = \left(\begin{array}{cc}
A_- & B_-\\
A_+ & B_+\end{array}\right).\eqno(A.2)$$
With the Dirac matrix conventions of footnote (1), we find that (A.1) leads to
the following consistency conditions upon identifying corresponding
terms dependent on $\lambda$, $\partial_t\lambda$ and $\partial_\rho\lambda$:
$$\left[C_\rho \left(\partial_t \pm \partial_p\right) \pm \frac{1}{2} S_\rho\right]
A_{\pm} = -i\Delta A_{\mp} \eqno(A.3a)$$
$$\left[C_p \left(\partial_t \pm \partial_\rho\right) \pm \frac{1}{2} S_\rho\right]
B_{\pm} = -i\Delta B_{\mp} \eqno(A.3b)$$
$$i C_\rho A_\pm = k\left[\left(-C_t S_\rho\right)A_{\mp} + 
\left(1 + S_\rho S_t\right)B_\mp \right]\eqno(A.4a)$$
$$i C_\rho B_\pm = k\left[\left(-1 +S_\rho S_t\right)A_{\mp} + 
\left(C_t S_\rho \right)B_\mp \right]\eqno(A.4b)$$
$$i A_\pm = \mp k\left[S_t A_\mp + C_t B_\mp \right]\eqno(A.5a)$$
$$i B_\pm = \mp k\left[C_t A_\mp - S_t B_\mp \right].\eqno(A.5b)$$
(Here, $C_p = \cos \rho$, $S_\rho = \sin_\rho$ etc., $\Delta = m - kM$ and
we have set $a = 1$.)

The algebraic eqs. (A.4 - A.5) are solved by
$$B_- = \tan \left(\frac{\rho - t}{2}\right) A_- = \frac{S_\rho - S_t}{C_\rho -
C_t} A_- \eqno(A.6a)$$
$$A_+ = \tan \left(\frac{\rho + t}{2}\right) B_+ \eqno(A.6b)$$
$$B_+ = \frac{i}{k}  \left(\frac{C_t + C_\rho}{S_{\rho -
t}}\right) B_- \eqno(A.6c)$$
and
$$k^2 =1 \; .\eqno(A.7)$$
The differential eqs. of (A.3) are consistent with (A.6) if
$\Delta = -k$, or in other words
$$m = k(M-1)\; .\eqno(A.8)$$
Finally, by using (A.6) and (A.3b) we find that $B_-$ satisfies the equations
$$\partial_- B_- = \frac{1}{\cos(x_+ - x_-)}\left[\frac{1}{2} \sin (x_+ - x_-)
+ \frac{\cos(x_+)}{\sin(x_-)}\right]B_-\eqno(A.9a)$$
and
$$\cos\left(x_+ - x_-\right)\partial_+ \left(
\frac{\cos(x_+)}{\sin(x_-)} B_-\right)
= -\frac{1}{2}\left[\frac{1}{2}\sin (x_+ - x_-)
\left(\frac{\cos(x_+)}{\sin(x_-)}\right) +1\right]B_-\eqno(A.9b)$$
where
$x_\pm = \frac{1}{2}(t \pm \rho )$. Eq. (A.9) serves to fix the
dependence of $B_-$ on $x_+$ and $x_-$; from eq. (A.6) we then find
$B_+$, $A_+$ and $A_-$.  A linear relationship between $\chi$ and $\lambda$
has been established.

\end{document}